\documentclass[10pt,conference]{IEEEtran} 
\usepackage{cite}
\usepackage{amsmath,amssymb,amsfonts}
\usepackage{algorithmic}
\usepackage{graphicx}
\usepackage{textcomp}
\usepackage{xcolor}
\usepackage{comment}
\usepackage{tabularx}
\usepackage{makecell}
\usepackage{graphicx}
\usepackage{cellspace}
\usepackage{mdframed}
\usepackage{array}
\usepackage{makecell}
\usepackage{balance}
\usepackage{url}
\usepackage{tcolorbox}
\usepackage[raggedrightboxes]{ragged2e}

\mdfdefinestyle{MyFrame}{
    innertopmargin=4pt,
    innerbottommargin=4pt,
    innerrightmargin=4pt,
    innerleftmargin=4pt,
    backgroundcolor=gray!15!white
        }

\usepackage{xcolor}
\usepackage{ifthen}
\newboolean{commentEnabled}
\setboolean{commentEnabled}{true}
\newcommand{\zibran}[1]{\ifthenelse{\boolean{commentEnabled}}{\textcolor{red}{Zibran: #1}}{}}
\newcommand{\rabbi}[1]{\ifthenelse{\boolean{commentEnabled}}{\textcolor{blue}{Rabbi: #1}}{}}

\usepackage{todonotes}

\usepackage{enumitem}

\begin{document}
\bstctlcite{IEEE-NoDash:BSTcontrol} 



\title{Understanding Software Vulnerabilities in the Maven Ecosystem: Patterns, Timelines, and Risks}

\author{\IEEEauthorblockN{
Md Fazle Rabbi ~~~~~~~~ Rajshakhar Paul ~~~~~~~~ Arifa Islam Champa  ~~~~~~~~ Minhaz F. Zibran}
\IEEEauthorblockA{\textit{Department of Computer Science, Idaho State University, Pocatello, ID, United States} \\
\{mdfazlerabbi, rajshakharpaul, arifaislamchampa, zibran\}@isu.edu}
}

\maketitle

\begin{abstract}
Vulnerabilities in software libraries and reusable components cause major security challenges, particularly in dependency-heavy ecosystems such as Maven. This paper presents a large-scale analysis of vulnerabilities in the Maven ecosystem using the Goblin framework. Our analysis focuses on the aspects and implications of vulnerability types, documentation delays, and resolution timelines. We identify 77,393 vulnerable releases with 226 unique CWEs. On average, vulnerabilities take nearly half a decade to be documented and 4.4 years to be resolved, with some remaining unresolved for even over a decade. The delays in documenting and fixing vulnerabilities incur security risks for the library users emphasizing the need for more careful and efficient vulnerability management in the Maven ecosystem. 
\end{abstract}


\vspace{-0.1cm}
\section{Introduction}

Modern software applications heavily rely on libraries and external dependencies \cite{bauer2012structured}. These components help developers speed up progress and reduce costs by providing reusable code. Studies have found that third-party libraries often contribute between 30\% to 90\% of the source code in typical applications~\cite{bandharangshi2014, rivera2019}. However, while these libraries offer significant advantages, they also introduce substantial security risks \cite{plate2015impact}. Vulnerabilities in libraries can spread across software ecosystems, affecting millions of users. High-profile incidents, such as Log4Shell vulnerability in Apache Log4j~\cite{everson2022log4shell} and the SolarWinds supply chain attack~\cite{martinez2021software
}, demonstrate the serious impact of compromised dependencies.


The prevalence of software vulnerabilities has raised significant concerns about the security and reliability of modern software systems. As software ecosystems expand, their attack surfaces increase, making the security of external libraries a critical concern for developers and organizations. Recent data from the National Vulnerability Database (NVD) shows a sharp increase in vulnerabilities. Projections estimate 35,000 new vulnerabilities in 2024, marking a 25\% rise from the previous year~\cite{henriques2024}. Many of these vulnerabilities are found in widely used software libraries, pointing to increasing security risks in the software supply chain. If left unaddressed, these vulnerabilities can affect entire software ecosystems. 

The Common Weakness Enumeration (CWE) framework offers a structured approach for identifying and classifying software weaknesses \cite{christey2013common}, which are foundational issues in code that can lead to security vulnerabilities if exploited. While CWE helps developers and organizations better understand and mitigate potential security risks, significant knowledge gaps remain regarding how vulnerabilities develop within the libraries. For instance, vulnerabilities often remain undocumented for extended periods, leaving users unaware of the risks associated with the libraries they rely on. Even when vulnerabilities are disclosed, the process of patching them can vary significantly across projects, leading to prolonged exposure. Additionally, understanding the common patterns of software weaknesses, such as those categorized under CWE, is critical for prioritizing mitigation efforts.



Addressing these gaps is essential for improving security practices, particularly in ecosystems like Maven that rely heavily on dependencies. Vulnerabilities in libraries pose not only potential risks but also significant threats to applications across many industries. By better understanding the types of vulnerabilities, their discovery timelines, and the resolution processes, developers can make more informed decisions about which libraries to use, create more robust security policies, and enhance strategies for managing software supply chain risks.



Towards this goal, we, therefore, examine the key aspects of software vulnerabilities in Maven ecosystem by investigating the following three research questions (RQs):

\vspace{2pt}

\noindent \textit{\textbf{RQ1:}  What are the most frequently occurring software weaknesses in libraries?} 

    
\vspace{2pt}
\noindent \textit{\textbf{RQ2:} How long do vulnerabilities remain undocumented, and what percentage of releases are deployed with known vulnerabilities?} 


\vspace{2pt}
\noindent \textit{\textbf{RQ3:} What are the typical resolution times for vulnerabilities, and what proportion of libraries remain unresolved?}


\vspace{5pt}

Our findings provide actionable insights for improving software security practices, particularly in dependency-heavy ecosystems like Maven. We make our scripts and data publicly available as a replication package~\cite{replicationPackage}.

\vspace{-0.1cm}
\section{Methodology}

In this study, we use the Goblin framework~\cite{goblin} as the primary data source, which provides a Neo4j-based dependency graph of the Maven Central repository. The dataset includes 658,078 artifact or library nodes and 14,459,139 release nodes, where each library node has a unique Maven identifier and release nodes capture version details and timestamps. 
Additionally, the dataset version \texttt{with\_metrics\_goblin\_maven\_30\_08\_24.dump}~\cite{jaime2024goblin}, dated August 30, 2024, includes 44,035,495 \textit{AddedValue} nodes. These AddedValue nodes include enriched metrics such as Common Vulnerabilities and Exposures (CVEs) sourced from the OSV dataset~\cite{osv2024}, along with other attributes like freshness, popularity, and speed.

From the dataset, we identify 77,393 releases with vulnerabilities. These releases are associated with 1,411 unique libraries. We then retrieve all releases from these 1,411 libraries, regardless of whether they have vulnerabilities. This results in a refined dataset comprising 125,816 releases associated with the 1,411 libraries. This \textit{refined dataset} (RD) serves as the basis for further analysis in our study.

\subsection{Identifying Frequent CWEs} 

To address RQ1, we analyze the structure of CVE information in RD. Each CVE entry comprises three elements: \textit{cwe}, \textit{severity}, and \textit{name}. CWE provides a standardized classification of software weaknesses. Severity indicates the criticality of the vulnerability and is categorized as \textit{critical}, \textit{high}, \textit{medium}, or \textit{low}.  The CVE name is a unique identifier for publicly disclosed vulnerabilities.

We extract the CWE information from the CVE data and compute the frequency of each unique CWE. The frequencies are then ranked in descending order to identify the most common weaknesses. This ranking provides insights into the dominant security weaknesses in Maven libraries.

\subsection{Analyzing Documentation Delays and Vulnerabilities}
\label{Method_RQ2}

We consider a vulnerability `documented' when it is officially assigned a CVE identifier and its details are made publicly available in a vulnerability database, such as the NVD. The time to documentation is the period between the first appearance of a vulnerability in the release of a library and the date when the vulnerability is officially published with a CVE identifier in a vulnerability database.
To determine how long vulnerabilities remain undocumented and the proportion of releases deployed with known vulnerabilities, we follow these steps:

\subsubsection{Data Preprocessing} 
We convert the release timestamps in RD from Unix format to a human-readable date-time format using Python’s \texttt{datetime} module.


\subsubsection{Fetching CVE Publication Dates} 
We extract CVE names from RD and retrieve their publication dates from the NVD through web scraping. The URL used for this process is: \url{https://nvd.nist.gov/vuln/detail/{cve_name}}, where \{cve\_name\} is dynamically replaced with CVE names. Some CVEs in the dataset are listed as \textit{UNKNOWN}, making it infeasible to fetch their publication dates. To maintain data quality, we exclude these entries, reducing the dataset from 1,411 to 1,381 libraries. Note that we exclude \textit{UNKNOWN} CVEs only from the RQ2 analysis and not from other parts of this study.



\subsubsection{Calculating Documentation Delays}

For each library, we calculate the time difference between the publication date of a CVE and the release date of the library version where the CVE first appears.


\begin{itemize}
    \item Positive time differences indicate delays in documenting vulnerabilities after a release.
    \item Negative time differences reveal releases deployed with known vulnerabilities.
\end{itemize}


\subsection{Calculating Fixing Timelines and Unfixed Vulnerabilities} 
To address RQ3, we analyze how long it takes to resolve vulnerabilities and quantify unresolved vulnerabilities:

\subsubsection{Tracking CVE Fixes} 
For each library in RD, we identify the first release where a CVE appears and the subsequent release where it is resolved. The time to fix the CVE is then calculated as the difference between the release dates of these two versions.

\subsubsection{Handling Ambiguities} 
We identify cases where multiple identical CVE names appear within the same release. For instance, the release \texttt{org.webjars.npm:jquery:1.11.3} contains two `UNKNOWN' CVE entries. Such occurrences introduce ambiguity when later releases fix an `UNKNOWN' CVE, as it is unclear which specific CVE has been resolved. To address this ambiguity, we create a unique identifier by merging the CVE name with its corresponding CWE classification. This approach ensures that each vulnerability is uniquely tracked across releases. After merging, we verify that no duplicate entries remain, which allows for accurate tracking of vulnerabilities.



\subsubsection{Unresolved Vulnerabilities}
We track CVEs that remain unresolved in the latest releases of each library. For these CVEs, we calculate the time difference between their first appearance in a release and Goblin's most recent update, which is August 30, 2024.

\section{Analysis and Findings}

\subsection{Frequency of CWE Types in Libraries}

In RD, we have 77,393 vulnerable releases from 1,411 unique libraries. 
These represent only 0.05\% of all releases, indicating that 99.5\% of the releases are non-vulnerable.

\begin{figure}[htbp]
    \centering
    \includegraphics[width=0.48\textwidth]{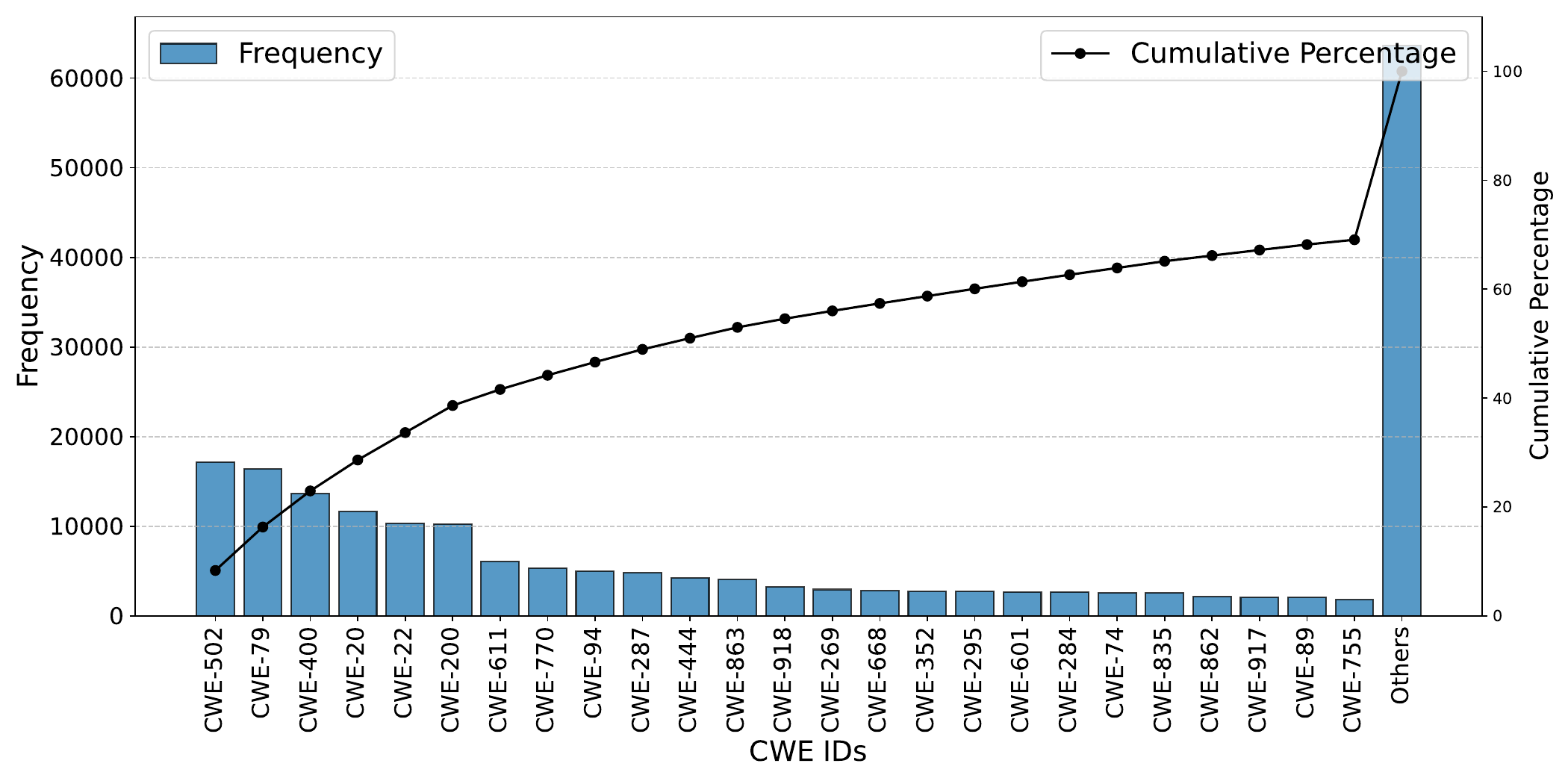} 
    \vspace{-0.2cm}
    \caption{Frequency of the top 25 CWEs and their cumulative percentages}
    \label{fig:pareto_cwe}
\end{figure}

From the vulnerable releases, we identify 226 unique CWEs. To illustrate the prevalence of these CWEs, we use a Pareto chart in Figure~\ref{fig:pareto_cwe} to display the 25 most frequently observed CWEs. 
The chart also shows their contribution to the cumulative percentage of occurrences. 
Each CWE is represented as a bar showing its frequency, with an additional `Others' category that aggregates the less frequent CWEs. 



The most frequently observed CWE is CWE-502 (Deserialization of Untrusted Data), with 17,187 occurrences. It is followed closely by CWE-79 (Cross-site Scripting), which has 16,426 occurrences. Other significant CWEs include CWE-400 (Uncontrolled Resource Consumption), CWE-20 (Improper Input Validation), CWE-22 (Path Traversal), and CWE-200 (Information Exposure). Each of these six CWEs has a frequency exceeding 10,000.

The cumulative frequency of the top 25 CWEs amounts to 142,166 occurrences. This accounts for 69.07\% of the total 205,822 CWE occurrences in the dataset. This indicates that a small subset of CWEs is responsible for the majority of vulnerabilities, with the remaining 201 CWEs contributing only 30.93\% of the total occurrences. 


Our results also indicate that a significant portion of top 25 weaknesses are related to \textit{input handling and validation} issues (CWE ID: 502, 79, 20, 22, 94, 611, 918, and 89). Another prevalent category belongs to \textit{access control and authorization} related issues (CWE ID: 287, 668, 284, 601, and 269).



\vspace{-0.1cm}
\begin{tcolorbox}[boxrule=0.5pt, boxsep=-2pt, left=5pt, right=5pt]
\textbf{Finding 1:} 
\textit{A few CWEs dominate the distribution of vulnerabilities in libraries. The top 25 CWEs account for 69.07\% of all occurrences, with CWE-502 and CWE-79 being the most frequent. Furthermore, a substantial portion of these weaknesses are related to \textit{``input handling and validation"} and \textit{``access control and authorization"} issues, underscoring key areas for mitigation efforts.}
\end{tcolorbox}

\subsection{Time to Documentation and Known Vulnerability Incidence}

We analyze the dataset of 1,381 libraries and identify a total of 3,407 CVEs after processing the data as described in Section~\ref{Method_RQ2}. For each CVE, we calculate the time difference as the number of days between the release date of the library and the publication date of that CVE.  Further analysis reveals that some CVEs appear multiple times across different libraries.  We find 2,822 unique CVEs, with 328 appearing in multiple libraries. To avoid duplication, we retain only the earliest occurrence of each CVE across libraries, ensuring each is counted once. 


Out of all the releases, we find that only ten releases are deployed with known vulnerabilities, as indicated by negative time differences. This small fraction implies that the deployment of releases with known vulnerabilities is an extremely rare event.

\vspace{-0.1cm}
\begin{figure}[htbp]
    \centering
    \includegraphics[width=0.48\textwidth]{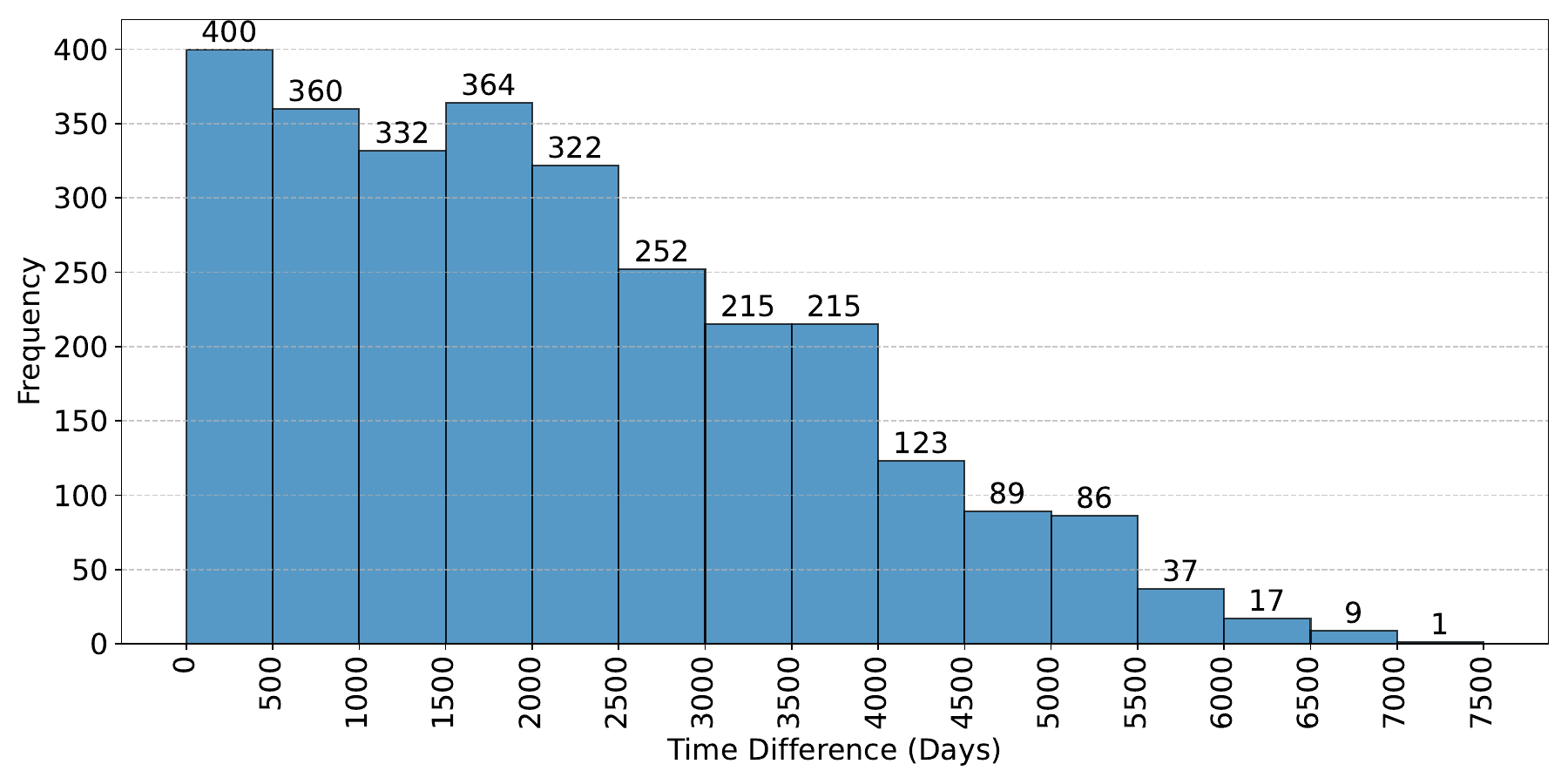} 
    \vspace{-0.2cm}
    \caption{Time taken to document vulnerabilities after library release}
    \label{fig:documented_time}
\end{figure}

The majority of the vulnerabilities, however, are documented after the release date. We focus on the positive time differences, which represent the time it takes for vulnerabilities to be formally documented after a release. Figure~\ref{fig:documented_time} illustrates the frequency distribution of these documentation delays. Although 400 ($\approx 14.2\%$) CVEs are documented within the first 500 days, a significant portion-- 1,378 out of 2,822 CVEs ($\approx$ 48.8\%) takes between 500 and 2,500 days to be documented. This delay highlights a substantial lag in the formal documentation of vulnerabilities, potentially leading to delayed awareness about the vulnerability, prolonged vulnerability exposure, and hindered mitigation efforts.


The mean documentation delay across all CVEs is 2,173.76 days, or roughly 5.95 years, with a minimum of just 1 day and a maximum of 7,078 days. The standard deviation of 1,502.36 days highlights the variability in documentation times.  While some vulnerabilities are addressed relatively quickly, many others remain unrecognized for an extraordinarily long time before being formally documented.


\vspace{-0.1cm}
\begin{tcolorbox}[boxrule=0.5pt, boxsep=-2pt, left=5pt, right=5pt]
\textbf{Finding 2:} \textit{The deployment of releases with known vulnerabilities is exceedingly rare. However, vulnerabilities take a long time to be documented, with an average delay of almost six years.}
\end{tcolorbox}

\subsection{Resolution Durations and Unfixed CVEs}

For each CVE observed in the releases of 1,411 libraries, there are two possible outcomes: the CVE is either resolved in subsequent releases or remains unresolved in the latest release of the library. Our analysis identifies a total of 3,518 CVEs across these libraries. Of those, 3,177 CVEs (90.3\%) are resolved in later releases, while 341 CVEs (9.7\%) remain unresolved in the most recent releases. Notably, 231 ($\approx 16.4\%$) libraries have unresolved CVEs in their latest release.

\vspace{-0.1cm}
\begin{figure}[htbp]
    \centering
    \includegraphics[width=0.48\textwidth]{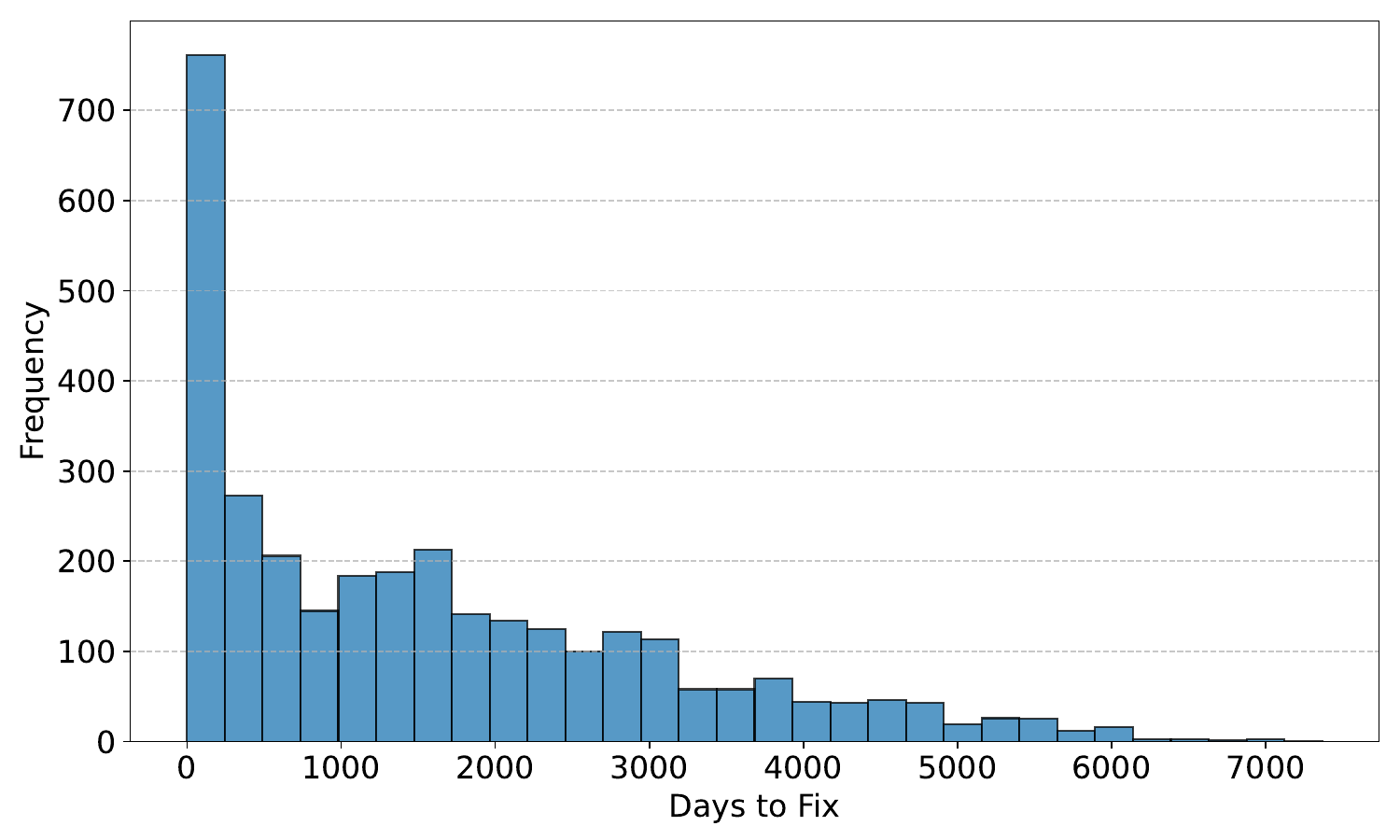}
    \vspace{-0.2cm}
    \caption{Frequency distribution of the time required to fix CVEs}
    \label{fig:fixed_time}
\end{figure}
\vspace{-0.15cm}

Figure~\ref{fig:fixed_time} presents the distribution of the time taken to resolve CVEs across all libraries. The x-axis represents the number of days required to fix vulnerabilities. It is divided into 30 bins, each covering approximately 246 days, with the bin size determined by the data distribution. The y-axis represents the frequency of CVEs resolved within these timeframes. Out of all CVEs, 761 (23.95\%) are resolved within the first 246 days. However, the remaining CVEs exhibit a long tail in the distribution, which indicates that many vulnerabilities take significantly longer to resolve.


On average, it takes 1,588.70 days ($\approx$ 4.4 years) to fix a CVE across all libraries. The resolution times vary widely, ranging from 0 days to a maximum of 7,367 days, with a standard deviation of 1,495.56 days.

\begin{figure}[htbp]
    \centering
    \includegraphics[width=0.48\textwidth]{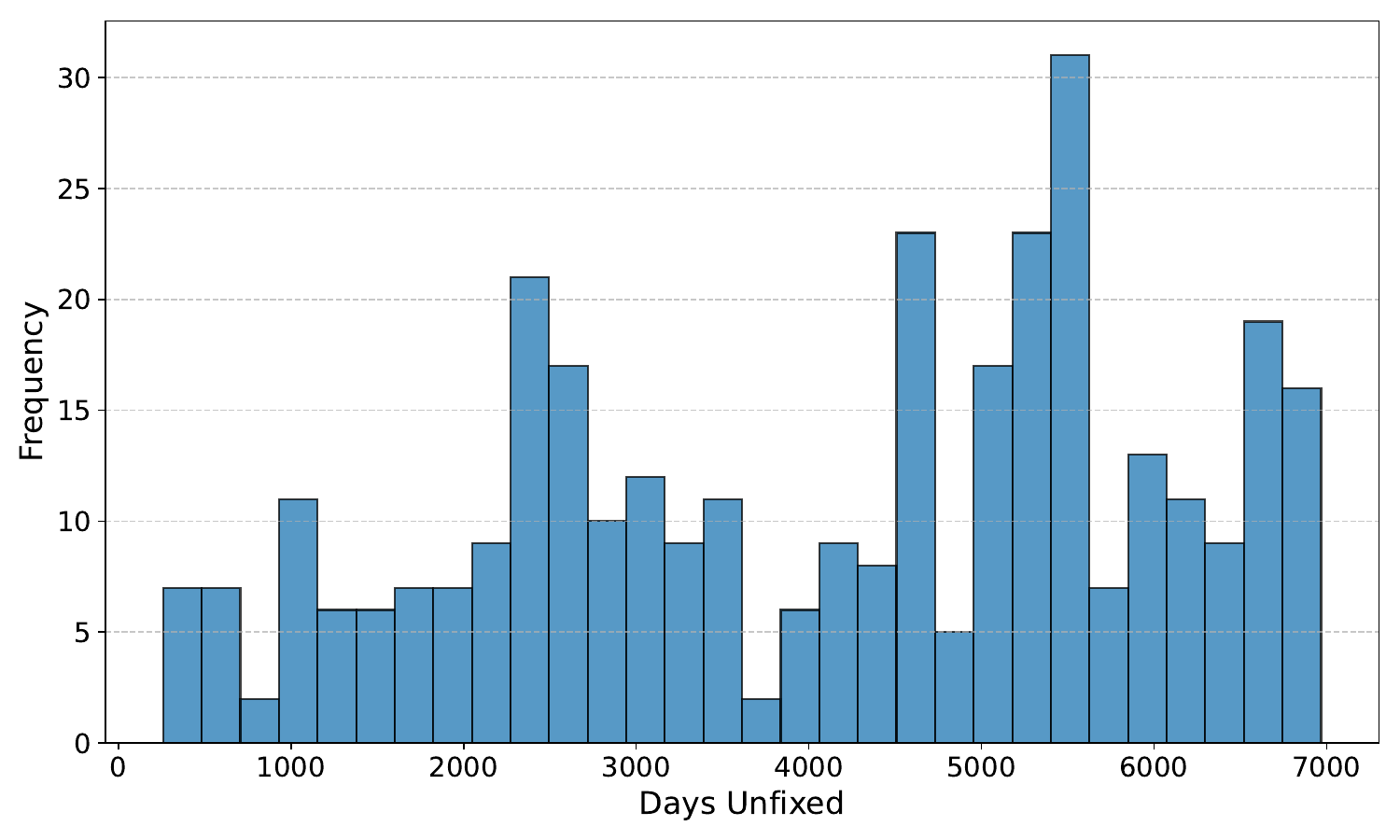} \vspace{-0.3cm}
    \caption{Unresolved CVEs by the number of days remained unfixed}
    \label{fig:unfixed_time}
\end{figure}

For unresolved CVEs, Figure~\ref{fig:unfixed_time} depicts the distribution of their duration across all libraries. The x-axis shows the number of days these vulnerabilities have remained unresolved, divided into 30 bins, with each bin representing $\approx$ 224 days. The y-axis represents their frequency. The data reveals that only 52 CVEs (15.25\%) have been unresolved for less than 2,000 days. The majority of unresolved CVEs, 144 (42.23\%), fall within the range of 2,000 to 5,000 days, while 145 CVEs (42.52\%) have remained unresolved for over 5,000 days.

\vspace{-0.15cm}

We find that these unresolved vulnerabilities have existed for an average of 4,118.38 days ($\approx$ 11.3 years), significantly longer than the time typically required to fix resolved CVEs. The duration of unresolved CVEs ranges from 259 days to 6,968 days, with a standard deviation of 1,856.74 days. 


\vspace{-0.1cm}
\begin{tcolorbox}[boxrule=0.5pt, boxsep=-2pt, left=5pt, right=5pt]
\textbf{Finding 3:} \textit{The majority of CVEs (90.3\%) are resolved in subsequent releases, but the resolution process often takes several years. Additionally, 9.7\% of CVEs remain unresolved in the latest releases of libraries, with some remaining unfixed for over a decade.}
\end{tcolorbox}

\section{Threats to validity}\label{sec:threats}
A potential limitation of this work may be the removal of `UNKNOWN' CVEs because they lack publication dates, which leads to the removal of 30 libraries from the dataset used in the RQ2 analysis. However, this has minimal impact, as we still analyze 1,381 ($\approx$ 97.9\% of total) vulnerable libraries, which is sufficient to draw reliable conclusions about delays in vulnerability documentation.

For RQ3 analysis, we consider the dataset's latest release date, August 30, 2024, as the current date to calculate unresolved vulnerability durations. It is important to note that as time progresses, the status of vulnerabilities may change, and some unresolved vulnerabilities identified in this study may now be resolved or updated in subsequent dataset releases.

We focus on libraries from the Maven Central, using the Goblin framework for data collection and analysis. While Maven Central is a widely used, the findings may not fully apply to ecosystems like PyPI, or npm, where dependency management and vulnerability reporting practices may differ. 

\section{Related Work}
Prior work examined software vulnerabilities across different package ecosystems~\cite{rabbi2024ai, Shafin2025faster, Rabbi2025understanding, Rabbi2025chasing}. In Python, Alfadel et al.~\cite{alfadel2023empirical} found Cross-Site Scripting (XSS) as the most frequent vulnerability, with a median discovery time of three years and a fix time of two months. Zerouali et al.~\cite{zerouali2022impact} studied npm and RubyGems, where Directory Traversal and XSS were common, with median discovery and fix times of 26 months and 11 days in npm. In Node.js, Alfadel et al.~\cite{alfadel2023discoverability} identified Prototype Pollution and SQL Injection as the most common vulnerabilities, with five CWE types accounting for 77.98\% of affected projects.

In the Rust ecosystem, Zheng et al.~\cite{zheng2023closer} found that vulnerabilities took a median of 1.9 years to be publicly disclosed, with many delays exceeding two years. Harzevili et al.~\cite{harzevili2023characterizing} found Memory Leaks (CWE-401) prevalent in machine learning libraries. In the Maven ecosystem, Mir et al.~\cite{mir2023effect} identified 114 distinct CWEs and found CWE-502 and CWE-79 as the most common, similar to our findings. Five CWEs accounted for 37\% of all discovered vulnerabilities.

While these studies provide valuable insights into vulnerability types and documentation delays, they often overlook unresolved vulnerabilities and their persistence across large-scale ecosystems. Our work fills this gap with a comprehensive analysis of Maven Central. We identify 226 CWEs across 77,393 vulnerable releases, with CWE-502 and CWE-79 being the most common. Our findings align with other ecosystems and extend the literature by quantifying documentation delays (5.95 years) and resolution timelines (4.4 years), with some vulnerabilities unresolved for over a decade.

\vspace{-0.1cm}
\section{Conclusion}

In this study, we have conducted a detailed investigation into software vulnerabilities within libraries from the Maven Central repository. Our analysis covers most frequent CWE types, documentation delays, and resolution timelines. We find that a small subset of CWEs dominate the distribution of vulnerabilities where 25 CWEs account for nearly 70\% of all cases. A significant portion of these weaknesses are related to input validation and access control issues, highlighting key areas for targeted mitigation efforts. While deployment with known vulnerabilities is rare, vulnerabilities take an average of nearly six years to be documented. We also find that majority of the vulnerabilities are resolved in subsequent releases. However, the average resolution time exceeds four years. The prolonged documentation and resolution delays indicate a need for improved process to track and fix vulnerabilities efficiently. In future, we plan to investigate how transitive dependencies contribute to the propagation of vulnerabilities and assess the real-world impact of unresolved vulnerabilities on dependency chains and critical applications. 




\balance

\section*{Acknowledgement}
This work is supported in part by the ISU-CAES (Center for Advanced Energy Studies) Seed Grant at the Idaho State University, USA.

\bibliographystyle{IEEEtran}
\bibliography{gender,Bug, MSR25}

\end{document}